\documentclass[twocolumn,aps,a4paper,superscriptaddress]{revtex4-2}
\usepackage{kpfonts}
\usepackage{amsmath}
\usepackage{amssymb}
\usepackage{times}
\usepackage{graphicx}
\usepackage[colorlinks,linkcolor=blue,citecolor=blue,urlcolor=blue,breaklinks]{hyperref}
\usepackage[normalem]{ulem}

\def\a{\alpha}
\def\b{\beta}
\def\g{\gamma}

\def\d{\delta}

\def\e{\epsilon}

\def\th{\theta}

\def\l{\lambda}

\def\m{\mu}
\def\n{\nu}

\def\p{\pi}
\def\P{\Pi}

\def\w{\omega}

\def\blk{{\mathbf k}}

\def\blq{{\mathbf q}}
\def\blr{{\mathbf r}}

\def\callA{\mbox{$\mathcal{A}$}}

\def\callF{\mbox{$\mathcal{F}$}}
\def\callG{\mbox{$\mathcal{G}$}}

\def\callO{\mbox{$\mathcal{O}$}}

\def\bra{\langle}
\def\ket{\rangle}

\def\Im{{\rm Im}}

\def\1op{\hat{\mathbbm{1}}}
\def\nn{\nonumber}

\begin{document}
\title{Exact formula with two dynamically screened electron-phonon couplings for 
 positive phonon-linewidths approximations}
\author{Gianluca Stefanucci}
\affiliation{Dipartimento di Fisica, Universit{\`a} di Roma Tor Vergata, Via della Ricerca Scientifica 1,
00133 Rome, Italy}
\affiliation{INFN, Sezione di Roma Tor Vergata, Via della Ricerca Scientifica 1, 00133 Rome, Italy}
\author{Enrico Perfetto}
\affiliation{Dipartimento di Fisica, Universit{\`a} di Roma Tor Vergata, Via della Ricerca Scientifica 1,
00133 Rome, Italy}
\affiliation{INFN, Sezione di Roma Tor Vergata, Via della Ricerca Scientifica 1, 00133 Rome, Italy}

\begin{abstract}  	
In this paper, we present an exact formula for the phonon linewidths 
involving only dressed electron-phonon couplings and ensuring the 
positivity property. The formula is designed to account for 
both nonadiabatic and correlation effects, and it provides 
an alternative proof 
that Density Functional Perturbation Theory 
calculations of phonon linewidths are not affected by a 
double counting of screening diagrams. 
Furthermore, we extend the treatment to nonequilibrium scenarios 
and offer a rigorous justification for employing the phononic 
Boltzmann equation. 
\end{abstract}

\maketitle

\section{Introduction}
Progress in designing new materials to improve device performances 
hinges on understanding the quantum mechanical behavior of a 
macroscopic number of electrons and nuclei. Among the valuable 
concepts researchers use to characterize a material, the phonon 
linewidth holds a prominent place~\cite{ziman_electrons_1960,grimvall_the-electron-phonon_1981}. 
It influences the thermal 
conductivity, thermal expansion, specific heat, and it 
may also indicate the occurrence of a structural phase transition.  

The popular formula~\cite{allen_neutron_1972,grimvall_the-electron-phonon_1981}
for the linewidth $\g_{\a\blq}$ of a phonon mode 
$\a$ with momentum $\blq$ (atomic units are used throughout this work)
\begin{align}
\g_{\a\blq}=&2\p\sum_{\m\n}
\sum_{\blk}|g^{s}_{\m\n,\a}(\blk,\blq)|^{2}
(f_{\n\blk}-f_{\m\blk+\blq})
\nn\\
&\times \d(\e_{\m\blk+\blq}-\e_{\n\blk}-\w_{\a\blq})
\label{allen}
\end{align}
involves the electronic band dispersions $\e_{\m\blk}$ and 
occupations $f_{\m\blk}$, the phonon 
frequencies $\w_{\a\blq}$, and the {\em statically screened} electron-phonon 
($e$-$ph$) coupling $g^{s}_{\m\n,\a}(\blk,\blq)
=\bra \m\blk+\blq|g^{s}_{\a-\blq}(\hat{\blr})|\n\blk\ket$.
This coupling is today routinely calculated using 
Density Functional Perturbation 
Theory (DFPT)~\cite{baroni_green_1987,gonze_dielectric_1992,savrasov_linear_1992}.
Equation~(\ref{allen}) with DFPT $g^{s}$
has been heavily utilized for 
decades~\cite{butler_phonon_1979,bauer_electron-phonon_1998,shukla_phonon_2003,lazzeri_phonon_2006,giustino_electron-phonon_2007,heid_effect_2010}, 
and continues to stand as a cornerstone even today.
It can be readily derived by applying the Fermi golden rule to a
(model) Hamiltonian in which electrons and phonons interact through 
$g^{s}$. 

In this Letter we show how to systematically improve 
Eq.~(\ref{allen}) using Many Body Perturbation Theory (MBPT). 
We derive an exact formula
expressed solely in terms of the {\em dynamically} dressed $e$-$ph$ 
coupling, and
ensuring the positivity of the linewidth at any temperature and for 
any approximation 
to $g^{s}$. We further extend the approach to the nonequilibrium 
realm, thereby providing a pathway to include non-Markovian screening effects 
in the phonon dynamics.

\section{Phonon self-energy and  linewidths}

\begin{figure}[tbp]
    \centering
\includegraphics[width=0.45\textwidth]{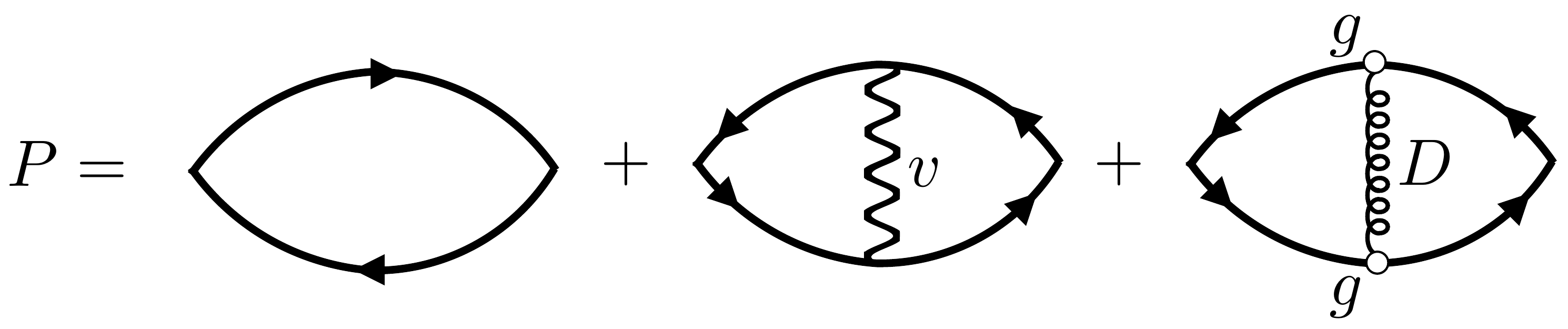}
\caption{Diagrammatic expansion of $P$ to lowest order in $g$ and 
$v$. Oriented lines represent electronic Green's functions and 
springs represent the phononic Green's function $D$.}
\label{Pol}
\end{figure}

In the first-principles 
Hamiltonian~\cite{marini_many-body_2015,stefanucci_in-and-out_2023}, 
electrons and phonons do not interact via $g^{s}_{\a\blq}(\blr)$, 
but rather through the {\em bare} 
$e$-$ph$ coupling $g_{\a\blq}(\blr)$. 
These couplings are related by the phonon-irreducible 
density response function $\tilde{\chi}(\blr,\blr';\w)$
(spatial convolutions are implicit)
\begin{align}
\tilde{\chi}(\w)=P(\w)
\big(1+v \tilde{\chi}(\w)\big)=\big(1+ \tilde{\chi}(\w) 
v\big) 
P(\w),
\label{chitilde}
\end{align}
where $P(\blr,\blr';\w)$ is the polarization and $v(\blr,\blr')$  is the bare Coulomb 
interaction.
In Fig.~\ref{Pol} we show the diagrammatic expansion 
of $P$ to second order in $g$ and first order in  $v$. 
The polarization is both $v$-irreducible and $D$-irreducible, $D$ 
being the phononic Green's function. The relation between 
$\tilde{\chi}$ (which is $v$-reducible and $D$-irreducible) and the 
full density response 
function $\chi$ (which is both $v$-reducible and $D$-reducible) is 
elaborated in the appendix material. Here, we limit ourselves to 
noting that $\chi$ and $\tilde{\chi}$ are identical only at clamped 
nuclei, i.e., for $g=0$.

Equation~(\ref{chitilde}) can be regarded as either a retarded ($R$)
or an advanced ($A$) equation (strictly at zero temperature 
it can also be regarded as a time-ordered equation). 
To distinguish between these cases, we use the superscripts 
$R$ or $A$ in the following discussion. The {\em 
dynamically screened} $e$-$ph$ coupling emerging from 
MBPT~\cite{feliciano_electron-phonon_2017,stefanucci_in-and-out_2023} reads 
\begin{align}
g^{s,R}_{\a\blq}(\w)\equiv \big(1+v\tilde{\chi}^{R}(\w)\big)\,g_{\a\blq}
=g_{\a\blq}+vP^{R}(\w)g^{s,R}_{\a\blq}(\w),
\label{gscreen}
\end{align}
and the like for the advanced coupling.
The  statically screened $g^{s}$ in Eq.~(\ref{allen}) 
is the zero-frequency value of Eq.~(\ref{gscreen}):
\begin{align}
g^{s}_{\a\blq}\equiv g^{s,R}_{\a\blq}(\w=0)=g^{s,A}_{\a\blq}(\w=0),
\label{ssg}
\end{align}
where in the second equality we use that
$\tilde{\chi}^{R}(\w=0)=\tilde{\chi}^{A}(\w=0)$~\cite{svl-book}.

MBPT leads to the 
following expression of the (retarded) phononic 
self-energy~\cite{feliciano_electron-phonon_2017,stefanucci_in-and-out_2023} 
\begin{align}
\P^{R}_{\a\b\blq}(\w)=g^{\ast}_{\a-\blq}P^{R}(\w)g^{s,R}_{\b-\blq}(\w)
=g^{s,A\ast}_{\a-\blq}(\w)P^{R}(\w)g_{\b-\blq},
\label{Piexact}
\end{align}  
featuring one bare and one screened $e$-$ph$ couplings --
the equivalence between the two expressions in Eq.~(\ref{Piexact})
follows from the 
properties 
$\tilde{\chi}^{A}(\blr,\blr';\w)=\tilde{\chi}^{R\ast}(\blr',\blr;\w)$, 
implying that 
\begin{align}
g^{s,A\ast}_{\a-\blq}(\w)=g^{\ast}_{\a-\blq}\big(1+\tilde{\chi}^{R}(\w)v\big).
\end{align}
As  $\g_{\a\blq}=-\Im[\P^{R}_{\a\a\blq}(\w_{\a\blq})]$, 
it would seem that the approximation in Eq.~(\ref{allen}) 
is affected by overscreening.
Here, by ``overscreening'', we refer to the conceptual error of double 
counting the screening diagrams.

A variational argument 
justifying Eq.~(\ref{allen}) from a first 
principles standpoint has been presented 
in Ref.~\cite{calandra_adiabatic_2010} and further elaborated in 
Ref.~\cite{berges_phonon_2023} in the context of 
linear-response 
time-dependent density functional theory (LR-TDDFT).
The idea is fully general and can be extended 
to MBPT to incorporate electron-phonon correlations in the 
polarization. To appreciate the difference with our 
derivation, we briefly illustrate how the variational argument
works in the MBPT context. Let $Q(\blr,\blr';\w)$ be an arbitrary function 
and consider the functional
\begin{widetext}
\begin{align}
\callF_{\a\b\blq}[Q(\w)]=g^{\ast}_{\a-\blq}\big(1+Q(\w)v\big)
P^{R}(\w)\big(1+vQ(\w)\big)g_{\b-\blq}
-g^{\ast}_{\a-\blq}Q(\w)vQ(\w)g_{\b-\blq}.
\label{variational}
\end{align}
It is a matter of simple algebra to verify that 
$\callF_{\a\b\blq}[\tilde{\chi}^{R}(\w)]=\P^{R}_{\a\b\blq}(\w)$, see 
Eq.~(\ref{Piexact}), and that $\callF_{\a\b\blq}[\tilde{\chi}^{R}(\w)+\d 
Q(\w)]=\P^{R}_{\a\b\blq}(\w)+\callO((\d Q)^{2})$.
Choosing $\d 
Q(\w)=\tilde{\chi}^{R}(0)-\tilde{\chi}^{R}(\w)$, 
the above properties imply  
\begin{align}
\P^{R}_{\a\b\blq}(\w)=g^{s\ast}_{\a-\blq}P^{R}(\w)g^{s}_{\b-\blq}
-g^{\ast}_{\a-\blq}\tilde{\chi}^{R}(0)v\tilde{\chi}^{R}(0)g_{\b-\blq}+\callO((\d Q)^{2})
\label{vararg}
\end{align}
\end{widetext}
\noindent
where we use Eq.~(\ref{ssg}). Taking into 
account that for $\a=\b$ the imaginary part of the second term 
vanishes, Eq.~(\ref{allen}) follows when the polarization is calculated 
to zeroth order in $v$ 
and $g$ (first diagram of 
Fig.~\ref{Pol}), and the correction $\callO((\d 
Q)^{2})$ is discarded.

We derive below an {\em exact} formula for the imaginary part of $\P^{R}$, 
and consequently for the phonon linewidth, which is manifestly positive 
definite and 
involves solely screened $e$-$ph$ couplings. 
Using Eqs.~(\ref{chitilde}) and (\ref{gscreen}), 
the phononic self-energy  can uniquely be written in terms of bare 
$e$-$ph$ couplings as
$\P^{R}_{\a\b\blq}(\w)=g^{\ast}_{\a-\blq}\tilde{\chi}^{R}(\w)g_{\b-\blq}$,
and the like for the advanced self-energy.
For retarded functions $F^{R}(\blr,\blr';\w)$, such as  $\tilde{\chi}^{R}$ 
or $P^{R}$, 
we define the spectral function 
\begin{align}
\callA_{F}(\blr,\blr';\w)\equiv 
i[F^R(\blr,\blr';\w)-F^A(\blr,\blr';\w)].
\end{align}
The functions $\callA_{\tilde{\chi}}$ and $\callA_{P}$
are related by (see below)
\begin{align}
\callA_{\tilde{\chi}}(\w)=\big(1+\tilde{\chi}^{R}(\w) 
v\big)\;\callA_{P}(\w)\;\big(1+v\tilde{\chi}^{A}(\w)\big).
\label{keyeq}
\end{align}
Therefore
\begin{align}
i\big[\P^{R}_{\a\b\blq}(\w)-\P^{A}_{\a\b\blq}(\w)\big]=
g^{s,A\ast}_{\a-\blq}(\w)\,\callA_{P}(\w)\,g^{s,A}_{\b-\blq}(\w),
\label{exactP<gd}
\end{align}
which is an exact relation at any temperature. For $\a=\b$ the lhs 
coincides with the imaginary part of the phononic self-energy. Explicitly 
writing down the 
spatial convolution, we obtain the first main result of this work
\begin{align}
-2\Im[\P^{R}_{\a\a\blq}(\w)]=
\int d\blr d\blr' g^{s,A\ast}_{\a-\blq}(\blr;\w)
\callA_{P}(\blr,\blr';\w)g^{s,A}_{\a-\blq}(\blr';\w).
\label{mainres}
\end{align}

Equation~(\ref{mainres}) clarifies the many-body approximations underlying 
Eq.~(\ref{allen}): 
By implementing the statically screened approximation, see 
Eq.~(\ref{ssg}),
and evaluating $P^{R}$ to zeroth order in $g$ and $v$ (first diagram of 
Fig.~\ref{Pol}), it is straightforward to verify that 
Eq.~(\ref{mainres}) with $\w=\w_{\a\blq}$ reduces to 
Eq.~(\ref{allen}). Of course, the statically screened 
linewidth in Eq.~(\ref{allen}) can be larger or smaller than the 
dynamically screened one in Eq.~(\ref{mainres}). 
However, any difference should be attributed to an underscreening or 
an overscreening of the statically screened 
approximation, not to the conceptual error of 
counting the same diagrams multiple times.
We also mention that Eq.~(\ref{allen})
can alternatively be derived under the mean field hypothesis that 
``electrons respond to the bare potential in the same way that
interacting electrons respond to the screened 
potential''~\cite{allenchapt}.

\begin{figure}[tbp]
    \centering
\includegraphics[width=0.45\textwidth]{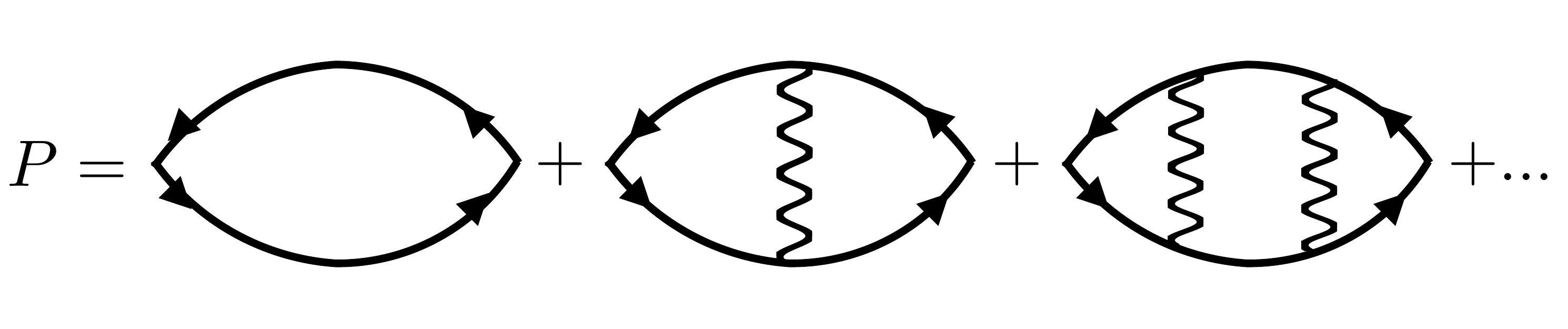}
\caption{T-matrix approximation of the polarization $P$.}
\label{PolTmat}
\end{figure}

A valuable feature of Eq.~(\ref{mainres}) is that it indicates 
how to go beyond Eq.~(\ref{allen}) 
via the inclusion of nonadiabatic 
effects in $g^{s,A}(\w)$ and correlation effects in $\callA_{P}(\w)$ without 
violating the positivity property. Specifically,
using for $\callA_{P}(\w)$
the diagrammatic expansion for positive spectra
developed in 
Refs.~\cite{stefanucci_diagrammatic_2014,uimonen_diagrammatic_2015},  
the function $-\Im[\P^{R}_{\a\a\blq}(\w)]$ is positive definite {\em for any} 
$g^{s,A}_{\a\blq}(\blr;\w)$,
and hence the linewidth $\g_{\a\blq}\geq 0$. For instance, the lowest order 
approximation of $P$ (first diagram of Fig.~\ref{Pol}) yields a positive 
$\callA_{P}(\w)$, and Eq.~(\ref{mainres}) provides the following 
generalization of Eq.~(\ref{allen}) 
\begin{align}
\g_{\a\blq}=&2\p\sum_{\m\n}
\sum_{\blk}|g^{s,A}_{\m\n,\a}(\blk,\blq;\w_{\a\blq})|^{2}
(f_{\n\blk}-f_{\m\blk+\blq})
\nn\\
&\times \d(\e_{\m\blk+\blq}-\e_{\n\blk}-\w_{\a\blq}),
\label{allengen}
\end{align}
where
\begin{align}
g^{s,A}_{\m\n,\a}(\blk,\blq;\w)=
\bra \m\blk+\blq|g^{s,A}_{\a-\blq}(\hat{\blr},\w)|\n\blk\ket.
\end{align}
Another example of positive definite approximation to $P$ is the 
T-matrix~\cite{uimonen_diagrammatic_2015}, see Fig.~\ref{PolTmat}. This 
polarization can be spectrally decomposed in terms of irreducible 
exciton-like energies 
$E^{\rm irr}_{\l\blq}$ and  wavefunctions $A^{{\rm 
irr},\l\blq}_{\m\n\blk}$, with normalization 
$s_{\l\blq}=\sum_{\m\n\blk}A^{{\rm irr},\l\blq\ast}_{\m\n\blk}
{\rm sgn}(f_{\n\blk}-f_{\m\blk+\blq})
A^{{\rm irr},\l\blq}_{\m\n\blk}=\pm 1$,~\cite{hannewald_theory_2000,perfetto_first-principles_2016,stefanucci_excitonic_2024}.
The T-matrix approximation leads
to the following generalization of Eq.~(\ref{allen})
\begin{align}
\g_{\a\blq}=&2\p\sum_{\l}s_{\l\blq}
|\callG^{A}_{\l,\a}(\blq;\w_{\a\blq})|^{2}
\d(E^{\rm irr}_{\l\blq}-\w_{\a\blq}),
\label{allenTmat}
\end{align}
where
\begin{align}
\callG^{A}_{\l,\a}(\blq;\w_{\a\blq})=\sum_{\m\n\blk}\sqrt{|f_{\n\blk}-f_{\m\blk+\blq}|}\;
A^{{\rm irr},\l\blq\ast}_{\m\n\blk}
g^{s,A}_{\m\n,\a}(\blk,\blq;\w)
\end{align}
can be interpreted as the dynamically screened
$e$-$ph$ coupling in the irreducible-exciton basis.

The take-home message is that 
computing the imaginary part first and 
then approximating $g^{s}$ 
is not equivalent to approximate $g^{s}$ 
first and then compute the imaginary part. In the latter case, 
for all $\w\neq 0$,
the two equivalent expressions of $\P^{R}$ in Eq.~(\ref{Piexact})
yield distinct results in the static limit, i.e., 
$g^{\ast}_{\a-\blq}P^{R}(\w)g^{s}_{\b-\blq}\neq 
g^{s\ast}_{\a-\blq}P^{R}(\w)g_{\b-\blq}$,
and neither guarantees 
the positivity of the linewidth. 
Going beyond the statically screened 
approximation, neither Eq.~(\ref{Piexact}) nor 
$\callF_{\a\b\blq}[\tilde{\chi}^{R}(\w)]$ in
Eq.~(\ref{variational}) guarantee that $\g_{\a\blq}\geq 0$ for a 
positive definite $\callA_{P}(\w)$. 
This also implies that only Eq.~(\ref{mainres}) is well suited for 
calculating the real part of the phononic self-energy via 
Kramers-Kronig relation, and thus the 
renormalized phonon frequencies $\w_{\a\blq}$~\cite{calandra_adiabatic_2010,lazzeri_nonadiabatic_2006,saitta_giant_2008} 
from the poles of the phononic Green's function.

It is worth remarking that the result in Eq.~(\ref{mainres}) does not 
justify the unconditional use of model Hamiltonians with a 
statically screened $e$-$ph$
coupling. Employing such Hamiltonians introduces an overscreening 
(to be understood in the original sense of multiple counting 
of the screening diagrams)
when the 
coupling to coherent 
phonons~\cite{stefanucci_semiconductor_2024,perfetto_theory_2024} or 
the
exciton-phonon dynamics~\cite{paleari_exciton-phonon_2022,stefanucci_excitonic_2024} are addressed, among 
other problems. Thus, although Eq.~(\ref{allen}) is not affected by 
overscreening, the 
original derivation, based on the Fermi golden rule, 
 is not correct.
It is also interesting to draw parallels between 
Eq.~(\ref{exactP<gd}) and the imaginary part of the screened 
Coulomb interaction $W=v+vPW=v+v\tilde{\chi}v$.
Multiplying Eq.~(\ref{keyeq}) on both the left and 
right by $v$, we obtain the well known result~\cite{svl-book} 
\begin{align}
\callA_{W}(\w)=W^{R}(\w)\callA_{P}(\w)W^{A}(\w),
\label{GWret}
\end{align}
which may (mistakenly) suggest the 
presence of overscreening here as well. 
The static approximation of 
Eq.~(\ref{GWret}), i.e.,  
$\callA_{W}(\w)\simeq W^{R}(0)\callA_{P}(\w)W^{A}(0)$
is at the basis of the formula for the electronic 
scattering rates within the GW 
approximation~\cite{marini_competition_2013,stefanucci_semiconductor_2024}, which are indeed free from 
overscreening.

\section{Nonequilibrium Phonons}

The entire formulation can be readily extended to systems out of 
equilibrium, thereby supporting the use of the Boltzmann 
equations~\cite{ziman_electrons_1960,ponce_toward_2018,sadasivam_theory_2017}, semiconductor 
Bloch equations~\cite{haug_quantum_1994,kira_many-body_2006}, 
and semiconductor electron-phonon equations (SEPE)~\cite{stefanucci_semiconductor_2024}
to explore the coupled dynamics of electrons and phonons. 
In Ref.~\cite{stefanucci_semiconductor_2024} we wrote about the 
phonon dynamics: 
``Currently, all nonequilibrium state-of-the-art 
methods dress {\em both} $g$'s, thereby suffering of a double 
counting problem.  The SEPE do not resolve this issue either.''
This is an erroneous statement as there is no issue to resolve, as demonstrated 
below.

Out of equilibrium, a time-ordered function 
$F^{T}$ and its retarded counterpart 
$F^{R}$
are independent of each other. It is therefore common to introduce  the so-called 
lesser/greater functions $F^{\lessgtr}(t,t')$, from which the
retarded function, $F^{R}(t,t')=\th(t,t')[F^{>}(t,t')-F^{<}(t,t')]$, 
as well as the time-ordered one 
follow~\cite{svl-book}. 
The Boltzmann equation for 
the phononic occupations reads
$\frac{d}{dt}f^{\rm ph}_{\a\blq}=S^{\rm ph}_{\a\blq}$~\cite{stefanucci_semiconductor_2024}, 
where the phononic scattering term  $S^{\rm ph}_{\a\blq}$ is a 
functional of the approximate self-energy
\begin{align}
\P^{\lessgtr}_{\a\b\blq}(t,t')=g^{s\ast}_{\a-\blq}P^{\lessgtr}(t,t')
g^{s}_{\b-\blq}.
\label{pi<>}
\end{align}
Let us show 
that $\P^{\lessgtr}$ in Eq.~(\ref{pi<>})
is not affected by overscreening.

We start from the exact relation 
\begin{align}
\P^{\lessgtr}_{\a\b\blq}(t,t')=g^{\ast}_{\a-\blq}\tilde{\chi}^{\lessgtr}(t,t')g_{\b-\blq}.
\label{pi<>ex}
\end{align}
To calculate 
$\tilde{\chi}^{\lessgtr}$ we use the Dyson equation on the Keldysh 
contour, $\tilde{\chi}=P+Pv\tilde{\chi}$,  
and the Langreth rules~\cite{svl-book,pavlyukh_photoinduced_2021} 
(space-time convolutions are implicit):
\begin{align}
\tilde{\chi}^{\lessgtr}=P^{\lessgtr}+P^{R}v\tilde{\chi}^{\lessgtr}+
P^{\lessgtr}v\tilde{\chi}^{A}.
\end{align}
Isolating $\tilde{\chi}^{\lessgtr}$ and taking into account that 
$(1+\tilde{\chi}^{R}v)(1-P^{R}v)=1$ we find
\begin{align}
\tilde{\chi}^{\lessgtr}=\big(1+\tilde{\chi}^{\rm R}
v\big)P^{\lessgtr}\big(1+v\tilde{\chi}^{\rm A}\big).
\end{align}
Inserting this result into Eq.~(\ref{pi<>ex}) we obtain the second 
main result of this work
\begin{align}
\P^{\lessgtr}_{\a\b\blq}(t,t')=\int d\bar{t}d\bar{t}'
g^{s,A\ast}_{\a-\blq}(t,\bar{t})P^{\lessgtr}(\bar{t},\bar{t}')g^{s,A}_{\b-\blq}(\bar{t}',t'),
\label{pi<>ex2}
\end{align}
where $g^{s,A}(t,t')$ is the (advanced) nonequilibrium screened $e$-$ph$ coupling. 
Implementing the static approximation at this stage, i.e.,  
$g^{s,A}_{\a\blq}(t,t')=\d(t,t')g^{s}_{\a\blq}$,
we retrieve Eq.~(\ref{pi<>}), which is therefore overscreening free.
The nonequilibrium results reduce to the previous ones for systems in 
equilibrium (or in a steady-state). In this case, all functions develop a dependence solely on 
the time difference and can be Fourier transformed. 
Taking into account that $
\P^{R}-\P^{A}=\P^{>}-\P^{<}$, and the like for $P$, we 
see that Eq.~(\ref{pi<>ex2}) reduces to Eq.~(\ref{exactP<gd}).

\section{Conclusions}

In conclusion, we have derived an exact formula for the phonon 
linewidths which involves only dressed $e$-$ph$ coupling. 
This finding confirms  
that the ``standard'' method of 
calculating phonon linewidths is not affected by the 
conceptual error of double counting the screening 
diagrams. In addition, it provides a 
pathway to incorporate nonadiabatic effects 
in $g^{s}$
and correlation effects in $P$ without violating the 
positivity property of the linewidth. 
The nonequilibrium extension also substantiates 
research on electron-phonon dynamics based on  
Boltzmann-like equations, and indicates how to improve the  
Boltzmann description. In fact,
the time-linear formulation of 
the Kadanoff-Baym equations for 
electrons~\cite{schlunzen_achieving_2020,perfetto_real_2022}
and phonons~\cite{karlsson_fast_2021} remains applicable 
when $g^{s}$ 
in Eq.~(\ref{pi<>ex2}) is screened at the RPA level, enabling the inclusion of 
non-Markovian effects.

We acknowledge funding from Ministero Universit\`a e 
Ricerca PRIN under grant agreement No. 2022WZ8LME, 
from INFN through project TIME2QUEST, 
from European Research Council MSCA-ITN TIMES under grant agreement 101118915, 
and from Tor Vergata University through project TESLA.
We also acknowledge useful discussions with Francesco Mauri, Samuel 
Ponce’, and Andrea Marini. 

\appendix

\section{Density response function}

The density response function $\chi(\blr,\blr';\w)$ 
for a system of electrons {\em and} phonons satisfies the Dyson equation
(spatial convolutions and summations over repeated indices are 
implicit)~\cite{stefanucci_in-and-out_2023,stefanucci_excitonic_2024}
\begin{align}
\chi(\w) 
=P(\w)+P(\w)\big(
v+g^{\ast}_{\blq\a}D^{0}_{\blq\a\b}(\w)g_{\blq\b}\big)\chi(\w),
\label{chi1}
\end{align}
where $D^{0}$ is the bare phonon 
Green's function. 
In accordance with the definition in Ref.~\cite{stefanucci_in-and-out_2023}, 
$D^{0}$ is the phonon Green's function of a system with frozen 
electronic density. As $D^{0}$ 
neglects the electronic feedback on the nuclei,
it does not exhibit poles at the physical phonon frequencies.  
Such definition allows for performing diagrammatic expansions and 
resummations without separating the phononic self-energy into 
adiabatic and nonadiabatic contributions.    

Let us consider the retarded version of Eq.~(\ref{chi1}). 
Iterating and grouping terms with the same powers of 
$g$ we find 
\begin{align}
\chi^{R}&=\tilde{\chi}^{R}+
\tilde{\chi}^{R} 
\big[g^{\ast}_{\a\blq}D^{0,R}_{\a\b\blq}g_{\b\blq}\big]
\tilde{\chi}^{R}
\nn\\
&+\tilde{\chi}^{R} \big[g^{\ast}_{\a\blq}D^{0,R}_{\a\b\blq}g_{\b\blq}\big]
\tilde{\chi}^{R}\big[g^{\ast}_{\a'\blq}D^{0,R}_{\a'\b'\blq}\
g_{\b'\blq}\big]\tilde{\chi}^{R} +\ldots,
\label{iter1}
\end{align}
where we recognize the $D$-irreducible density response function 
$\tilde{\chi}$ introduced in Eq.~(\ref{chitilde}). 
In Eq.~(\ref{iter1}) we take into account that for lattice 
periodic systems the double 
spatial convolution 
$g_{\b\blq}\tilde{\chi}^{R}g^{\ast}_{\a'\blq'}\propto \d_{\blq\blq'}$.
Using the definition in Eq.~(\ref{gscreen}), we can rewrite Eq.~(\ref{chi1}) in the following form
\begin{align}
\chi^{R}&= \tilde{\chi}^{R} + P^{R} \big[g^{s,R}_{\a-\blq} D^{0,R}_{\blq\a\b}
g^{s,A\ast}_{\b-\blq}\big]P^{R}
\nn\\
&+ P^{R} \big[g^{s,R}_{\a-\blq} D^{0,R}_{\a\b\blq}
g^{s,A\ast}_{\b-\blq}\big]P^{R} \big[g_{\a'-\blq} 
D^{0,R}_{\a'\b'\blq}
g^{s,A\ast}_{\b'-\blq}\big]P^{R}+\ldots,
\label{iter2}
\end{align}
where we use that $\tilde{\chi}^{A}=[\tilde{\chi}^{R}]^{\dag}$ and 
$g_{\a\blq}(\blr)=g^{\ast}_{\a-\blq}(\blr)$. 
In Eq.~(\ref{iter2}) we recognize the phononic self-energy of 
Eq.~(\ref{Piexact}).
Taking into account that the dressed phonon Green's function 
satisfies the Dyson equation 
$D^{R}_{\a\b\blq}=D^{0,R}_{\a\b\blq}+
D^{0,R}_{\a\a'\blq}\Pi^{R}_{\a'\b'\blq}D^{R}_{\b'\b\blq}$, we conclude that
\begin{align}
\chi^{R}(\w)&= \tilde{\chi}^{R}(\w) + 
P^{R}(\w)g^{s,R}_{\a-\blq}(\w) D^{R}_{\blq\a\b}(\w)
g^{s,A\ast}_{\b-\blq}(\w)P^{R}(\w),
\label{exactchiR}
\end{align}
which is an exact rewriting of the density response function.

Interestingly, equation~(\ref{exactchiR}) provides a direct access to the effective 
electron-electron ($e$-$e$) interaction. The $D$-irreducible density response in Eq.~(\ref{chitilde})
can be written in terms of the screened Coulomb interaction  
according to 
$\tilde{\chi}^{R}=P^{R}+P^{R}W^{R}P^{R}$. Inserting this result into 
Eq.~(\ref{exactchiR}), we find 
\begin{align}
\chi^{R}(\w)=P^{R}(\w)+P^{R}(\w)W_{\rm 
eff}^{R}(\w)P^{R}(\w),
\end{align}
where $W^{R}_{\rm eff}=W^{R}+
g^{s,R}_{\a-\blq} D^{R}_{\blq\a\b}
g^{s,A\ast}_{\b-\blq}$ is the effective $e$-$e$ interaction appearing in 
the Hedin-Baym equations~\cite{giustino_electron-phonon_2007,stefanucci_in-and-out_2023}.
This interaction has previously been inferred
through either  
field-theoretic 
treatments~\cite{vanleeuwen_first-principles_2004,giustino_electron-phonon_2007,marini_many-body_2015} 
or diagrammatic expansions~\cite{stefanucci_in-and-out_2023} of the 
electronic self-energy. Here, we have shown that $W_{\rm eff}$ naturally 
emerges also in the evaluation of the density response function.


%

\end{document}